 \documentclass{amsart}
\usepackage{amsmath,amssymb}

\newtheorem{theorem}{Theorem}[section]
\newtheorem{proposition}[theorem]{Proposition}
\newtheorem{lemma}[theorem]{Lemma}
\newtheorem{corollary}[theorem]{Corollary}
\newtheorem{conjecture}[theorem]{Conjecture}
\newtheorem{definition}[theorem]{Definition}

\numberwithin{equation}{section}

\def\DJ{{\hbox{{\thinspace}D\kern-.8em\raise.15ex\hbox{--}\kern.35em}}}
\def\DJo{{$\;$\kern-.4em{\DJ}okovi\'c}}

\def\NSERC{Supported in part by an NSERC Discovery Grant.}

\renewcommand{\subjclassname}{\textup{2000} Mathematics Subject
Classification}

\def\al{{\alpha}}
\def\be{{\beta}}

\def\sig{{\sigma}}

\def\vf{{\varphi}}
\def\la{{\lambda}}
\def\La{{\Lambda}}
\def\bR{{\mbox{\bf R}}}

\def\bC{{\mbox{\bf C}}}

\def\pH{{\mathcal{H}}}
\def\pM{{\mathcal{M}}}

\def\GL{{\rm GL}}
\def\Un{{\rm U}}

\def\tr{{\rm tr\;}}

\def\vek#1{|#1\rangle}
\def\kov#1{\langle#1|}

\begin{document}

\title[Generalized distillability conjecture]
{Generalized distillability conjecture and generalizations of
Cauchy-Bunyakovsky-Schwarz inequality and Lagrange identity}

\author {Dragomir \v{Z}. \DJo}
\address{Department of Pure Mathematics and Institute for Quantum Computing, University of Waterloo,
Waterloo, Ontario, N2L 3G1, Canada}
\email{djokovic@uwaterloo.ca}
\thanks{\NSERC}

\keywords{Distillability problem, bipartite entangled states, Werner states, hypermatrices, Cauchy-Bunyakovsky-Schwarz inequality, 
Lagrange identity}

\date{}

\begin{abstract}
The Quantum Information Theory is a reach source of fascinating 
problems in Linear and Multilinear Algebra. In this paper we 
discuss one of such problems, namely the Distillation Problem.

Let $\rho_k^W$, $k=1,2,\ldots,m$, be the critical Werner state in 
a bipartite $d_k\times d_k$ quantum system, i.e., the one that 
separates the 1-distillable Werner states from those that are 
1-indistillable. We propose a new conjecture (GDC) asserting that 
the tensor product of $\rho_k^W$ is 1-indistillable. This is much 
stronger than the familiar conjecture saying that a single 
critical Werner state is indistillable. We prove that GDC is true 
for arbitrary $m$ provided that $d_k>2$ for at most one index $k$. 
We reformulate GDC as an intriguing inequality for four arbitrary 
complex hypermatrices of type $d_1\times\cdots\times d_m$. 
This hypermatrix inequality is just the special case $n=2$ 
of a more general conjecture (CBS conjecture) for $2n$ arbitrary 
complex hypermatrices of the same type. Surprisingly, the case
$n=1$ turns out to be quite interesting as it provides hypermatrix 
generalization of the classical Lagrange identity.
We also formulate the integral version of the CBS conjecture and 
derive the integral version of the hypermatrix Lagrange identity.
\end{abstract}

\maketitle 
\subjclassname{ 15A69, 81P45 }
\vskip5mm

\section{Introduction} \label{Uvod}

There are two kinds of bipartite entangled quantum states. First,
the states which satisfy the Peres-Horodecki criterion of 
separability (i.e., having positive semidefinite partial transpose) 
are known as PPT entangled states. Second, those which violate 
this criterion are known as NPT entangled states.
The indistillable entangled states are also known as bound 
entangled. (See the next section for the definition of 
distillability and 1-distillability.) 
The problem of existence of bound NPT entangled states was
raised more than ten years ago \cite{MH1} and is still open. 
There are several papers \cite{DiV,WD,ML,LP,RV,DZ1} where evidence 
(numerical or theoretical) is provided for the existence of such 
states. Several researchers have proposed the conjecture 
\cite{MH1,DiV,WD} that such states exist. We refer to it as the 
Distillability Conjecture (DC).
This conjecture has been reduced \cite{MH1} to the case of Werner 
states. More precisely, for a fixed bipartite $d\times d$ quantum 
system, it is known that bound NPT entangled states exist if and 
only if (iff) a Werner state with the same properties exists.

The non-normalized Werner states in a $d\times d$ quantum system 
can be parametrized as $\rho^W(t)=1-tF$, $-1\le t\le1$, where $F$ 
is the usual flip operator (see its definition in the next section). 
They are separable for $t\le1/d$ and entangled for 
$t>1/d$ \cite{RW,WD}. They are 1-distillable for $t>1/2$ 
and 1-indistillable for $t\le1/2$ \cite{DiV,WD,ML}. Thus 
the state $\rho^W(1/2)$ separates the 1-distillable Werner states 
from those that are 1-indistillable. For this reason we refer to 
it as the critical Werner state. It has been shown that if the 
critical Werner state is indistillable, then this is also the case 
for all 1-indistillable Werner states \cite{DiV}. 

In Section \ref{GDC-hip} we propose much stronger conjecture 
which asserts that the tensor product of several critical Werner 
states $\rho_k^W=\rho_k^W(1/2)$, $k=1,\ldots,m$, acting in a 
$d_k\times d_k$ system (with arbitrary finite dimensions $d_k$) 
is indistillable. We refer to it as the Generalized Distillability 
Conjecture (GDC). We recall that there is an example \cite{PS} of a 
distillable state $\rho_{\bf Pyr}\otimes\rho^W(1/2)$ for two 
pairs of qutrits (i.e., in the case $d_1=d_2=3$), 
where $\rho_{\bf Pyr}$ is a particular PPT 
entangled state and, of course, $\rho^W(1/2)$ is the critical Werner 
state. In view of this example it seems rather foolhardy to propose 
a conjecture like GDC. However, subsequent reformulation of GDC and 
a further sweeping generalization, as well as analysis and resolution 
of some special cases led us to this proposal.

In Section \ref{Tenzor} we reformulate GDC as an inequality for
four complex hypermatrices of type $d_1\times\cdots\times d_m$.
Let $\pM_{\bf d}$ denote the space of such hypermatrices,
where ${\bf d}=(d_1,\ldots,d_m)$. We embed this GDC inequality 
into an infinite collection of similar inequalities 
\begin{equation} \label{CBS-nejed}
\Phi_{\bf d}^{(n)}(x^{(1)},\ldots,x^{(n)},u^{(1)},\ldots,u^{(n)})
\ge0, \quad m,n\ge1,
\end{equation} 
where $x^{(k)},u^{(k)}\in\pM_{\bf d}$ are arbitrary. The function 
$\Phi_{\bf d}^{(n)}$ is defined by the formula
\begin{eqnarray*} 
&& \Phi_{\bf d}^{(n)}(x^{(1)},\ldots,x^{(n)},
u^{(1)},\ldots,u^{(n)}) \\ \notag && \qquad 
= \sum_{Q\subseteq I_m} \left( \frac{-1}{n} \right)^{|Q|}
\sum_{i_p,j_p;\, p\in I_m\setminus Q} 
\quad \left| \sum_{i_q,j_q=i_q;\, q\in Q} \quad
\sum_{k=1}^n x_{\bf i}^{(k)}u_{\bf j}^{(k)} \right|^2,
\end{eqnarray*} 
where $I_m=\{1,2,\ldots,m\}$ and ${\bf i}=(i_1,i_2,\ldots,i_m)$
and ${\bf j}=(j_1,j_2,\ldots,j_m)$ with $i_k$ and $j_k$ 
running through $\{1,2,\ldots,d_k\}$.

We refer to the inequalities Eq. (\ref{CBS-nejed}) as the generalized 
Cauchy-Bunyakovsky-Schwarz (CBS) inequalites, and we conjecture 
that all of them are valid (CBS conjecture). The case $n=2$ 
corresponds to GDC. In the very first case, $m=n=1$, the inequality 
Eq. (\ref{CBS-nejed}) is just the classical CBS inequality.

In Section \ref{Osobine} we discuss some basic properties of the
function $\Phi_{\bf d}^{(n)}:\pM_{\bf d}^{2n}\to\bR$. In particular, 
we introduce natural actions of the unitary groups $\Un(d_k)$
and the general linear group $\GL_n(\bC)$ on $\pM_{\bf d}^{2n}$ 
and we show that $\Phi_{\bf d}^{(n)}$ remains invariant under 
these actions. Another property shows that the cases where some 
$d_k=1$ can be eliminated, i.e., the CBS conjecture can be reduced 
to the case where all $d_k>1$.

In Section \ref{Specijalan} we prove that the inequalities 
(\ref{CBS-nejed}) are true when $m=1$. We also show that they are 
true when $n=2$ and at most one $d_k>2$.

In Section \ref{Lagrange} we prove that the inequalities 
(\ref{CBS-nejed}) are true when $n=1$. Moreover, in that case we 
are able to express the function $\Phi_{\bf d}^{(1)}(x,u)$ as a sum 
of squares of real valued polynomials. This gives the hypermatrix 
generalization of the classical Lagrange identity, see 
Eq. (\ref{Lagr-alg}), valid over any commutative ring $\Lambda$.

As a concrete example, we write this identity in the case $m=2$, 
i.e., when $x$ and $u$ are ordinary $d_1\times d_2$ matrices with 
entries in $\Lambda$:
\begin{eqnarray*}
&& \left( \sum_{i=1}^{d_1} \sum_{j=1}^{d_2} x_{ij}^2 \right) \cdot
\left( \sum_{k=1}^{d_1} \sum_{l=1}^{d_2} u_{kl}^2 \right)
-\sum_{j=1}^{d_2} \sum_{l=1}^{d_2} \left( \sum_{i=1}^{d_1}
x_{ij}u_{il} \right)^2 \\
&& -\sum_{i=1}^{d_1} \sum_{k=1}^{d_1} \left( \sum_{j=1}^{d_2}
x_{ij}u_{kj} \right)^2 + \left( \sum_{i=1}^{d_1} \sum_{j=1}^{d_2}
x_{ij}u_{ij} \right)^2 \\
&& =\sum_{ \begin{array}{c} 1\le i<k\le d_1 \\ 1\le j<l\le d_2 
\end{array} } \left( x_{ij}u_{kl}-x_{il}u_{kj}-x_{kj}u_{il}
+x_{kl}u_{ij} \right)^2.
\end{eqnarray*}

In Section \ref{Integral} we formulate the integral version of the
CBS conjecture and derive the integral version of the hypermatrix 
Lagrange identity.

Finally, in Section \ref{Zakljucak} we summarize our results and 
conjectures. 

The symbols $*$, T and $\dag$ denote the complex conjugation, 
the transposition and the adjoint, respectively. 
For any positive integer $d$ we set $I_d=\{1,2,\ldots,d\}$.

\section{Preliminaries} \label{Prelim}

We consider a quantum system consisting of two parties, A and B (Alice and Bob), sharing a pair of particles. We denote by $\pH=\pH^A\otimes\pH^B$ the Hilbert space for this pair. We assume that $\pH^A$ and $\pH^B$ have the same finite dimension, $d\ge2$. A {\em product state} is a tensor product $\pi=\rho^A\otimes\rho^B$ of the states $\rho^A$ 
and $\rho^B$ of the first and second party, respectively. 

We remind the reader that a {\em state} or {\em density operator},
say $\rho^A$, is a positive semidefinite operator $\pH^A\to\pH^A$ 
with unit trace, $\tr \rho^A=1$. We often work with non-normalized 
states which are just nonzero positive semidefinite operators.
A {\em pure state} $\vek{\psi}\in\pH$ is a unit vector; two such 
states are considered the same if they differ only by a phase factor.
The density operator of the pure state $\vek{\psi}$ is the 
1-dimensional projector $\rho=\vek{\psi}\kov{\psi}$; it is 
independent of the choice of the phase factor of $\vek{\psi}$. 
We also refer to this $\rho$ as a {\em pure state}.

Any pure state $\vek{\psi}\in\pH$ can be written uniquely as 
$\vek{\psi}=\sum c_{ij}\vek{i,j}$, where $[c_{ij}]$ is a square 
matrix and we write $\vek{i,j}$ for $\vek{i}\otimes\vek{j}$.
The {\em Schmidt rank} of $\vek{\psi}$ is defined as 
the rank of the matrix $[c_{ij}]$. While this matrix depends on the 
choice of orthonormal (o.n.) bases of $\pH^A$ and $\pH^B$, its rank 
is independent of this choice. One can choose these o.n. bases so 
that $c_{ij}=0$ for $i\ne j$ and $c_{ii}\ge0$ for all $i$. 
Then the Schmidt rank of $\vek{\psi}$ is just the number of 
indexes $i$ such that $c_{ii}>0$. The pure states $\vek{\psi}$ 
of Schmidt rank 2 play an important role in this paper (see 
the definition of 1-distillability below).

We denote by $\rho^\Gamma$ the partial transpose $(T\otimes1)(\rho)$
of a density operator $\rho$ on $\pH$, where the transposition map 
$T$ is computed with respect to some o.n. basis of $\pH^A$. 

We shall use the following basic definitions.

\begin{definition} \label{Def-1}
Let $\rho$ be a density operator on $\pH=\pH^A\otimes\pH^B$.

(i) $\rho$ is {\em separable} if it can be written as a finite 
convex linear combination $\rho=\sum\lambda_i\pi_i$ 
$(\lambda_i\ge0,\,\sum\lambda_i=1)$ of product states $\pi_i$. 
It is {\em entangled} if it is not separable.

(ii) $\rho$ is {\em PPT} if its partial transpose $\rho^\Gamma$ 
is positive semidefinite. It is {\em NPT} if it is not PPT.

(iii) $\rho$ is $1$-{\em distillable} if 
$\kov{\psi}\rho^\Gamma\vek{\psi}<0$ for some $\vek{\psi}\in\pH$ 
of Schmidt rank $2$. It is $1$-{\em indistillable} if it is 
not $1$-distillable.

(iv) $\rho$ is {\em distillable} if $\rho^{\otimes m}$ is $1$-distillable for some $m\ge1$. It is {\em indistillable} if it is 
not distillable.
\end{definition}

For more background information on entanglement, separability and distillability of bipartite quantum states we refer the reader to one of the papers \cite{LC,DiV,WD,ML,JW} or the recent book \cite{BD}. 
It is well-known that the separable states are PPT \cite{AP}. It is 
also known that the distillable states are NPT \cite{MH2}. 
The question whether all NPT states are distillable has 
its origin in \cite{MH2}. We shall refer to this question as the 
{\em Distillability Problem}. The answer to this problem is 
affirmative for $d=2$, but  widely believed to be negative for 
$d\ge3$. Formally, the following conjecture, to which we refer as 
the {\em Distillability Conjecture} (DC), has been proposed in 
\cite[Sec. II]{DiV} (see also \cite{MH1,WD}).

\begin{conjecture} \label{DC}
There exist indistillable bipartite NPT states.
\end{conjecture}

Let us fix an o.n. basis $\vek{i}$, $i\in I_d$ of 
$\pH^A$, and an o.n. basis of $\pH^B$ for which we use the same notation. Due to the context, no confusion should arise. 

After fixing these bases, we can define the 
{\em flip operator} $F:\pH\to\pH$ by 
$$ F=\sum_{i,j} \vek{i,j} \kov{j,i}, $$
where the indexes $i,j$ run through $I_d$, and we use the common 
abbreviations $\vek{x,y}=\vek{x}\otimes\vek{y}$ and 
$\kov{x,y}=\kov{x}\otimes\kov{y}$.
The (non-normalized) Werner states on $\pH$ can be parametrized as
\begin{equation} \label{Werner-st}
\rho^W(t)=1-tF, \quad -1 \le t \le 1.
\end{equation}
Let $\vek{\vf}\in\pH$ be the maximally entangled (pure) state given by
$$ \vek{\vf}=\frac{1}{\sqrt{d}} \sum_{i} \vek{i,i}. $$
Its density matrix is the projector
$$ P=\frac{1}{d} \sum_{i,j} \vek{i,i} \kov{j,j}. $$
The partial transpose of $\rho^W(t)$ is $\sig^W(t)=1-tdP$.

The following facts about the Werner states are well-known.
\begin{proposition} \label{Werner-prop}
The Werner states $\rho^W(t)$ are: 

(a) separable iff $-1\le t\le 1/d$;

(b) NPT iff $1/d<t\le1$;

(c) $1$-distillable iff $1/2<t\le1$.
\end{proposition}
For (a) see \cite{RW,WD}, for (b) and (c) see \cite{DiV,WD,ML}. Since $\rho^W(1/2)$ separates the $1$-distillable Werner states from the 1-indistillable ones, we shall refer to it as the {\em critical Werner state}. The importance of Werner states for the distillability problem for bipartite states was first established in \cite{MH1}. 

\begin{proposition} \label{Horod-prop}
For a fixed $d\times d$ quantum system, 
DC is true iff there exists an indistillable NPT Werner state.
\end{proposition}

We end this section with two examples.

First, we verify that the Werner states $\rho^W(t)$ are 
$1$-distillable for $1/2<t\le1$. We take 
$\vek{\psi}=\vek{1,1}+\vek{2,2}$, which has Schmidt rank 2. Then 
$\langle \psi | \vf \rangle = 2/\sqrt{d}$ and
$P\vek{\psi}=(2/\sqrt{d})\vek{\vf}$. Hence
\begin{eqnarray*}
\kov{\psi}\sig^W(t)\vek{\psi} &=& \kov{\psi} 1-tdP \vek{\psi} \\
&=& \|\psi\|^2-td\kov{\psi} P \vek{\psi} \\
&=& 2-2t\sqrt{d} \langle \psi | \vf \rangle \\
&=& 2(1-2t) < 0.
\end{eqnarray*}

In our second example we consider the Hermitian operators 
$1-tdP$ where $t$ is a real parameter. Since its eigenvalues are
$1$ and $1-td$, it is positive semi-definite iff $t\le 1/d$.
These states are known in the literature as {\em isotropic states}
(see e.g. \cite{VW}).
The partial transpose of a separable state is also a separable 
state. Hence, the states $\sig^W(t)=1-tdP=\rho^W(t)^\Gamma$, 
$-1\le t\le 1/d$, are separable. We shall prove that the states
$1-tdP$, $t<-1$, are 1-distillable. According to 
Definition \ref{Def-1}, we have to show that 
$\kov{\psi} 1-tF \vek{\psi}<0$ for some vector $\vek{\psi}$ 
of Schmidt rank 2. The choice $\vek{\psi}=\vek{1,2}-\vek{2,1}$ 
works in all these cases. Indeed, we have 
$(1-tF)\vek{\psi}=(1+t)\vek{\psi}$ and so
$$ \kov{\psi} 1-tF \vek{\psi}=(1+t)\|\psi\|^2=2(1+t)<0. $$

\section{Generalized distillability conjecture} \label{GDC-hip}

We now assume that Alice and Bob share $m$ pairs of particles and
denote by $\pH_k=\pH_k^A\otimes\pH_k^B$ the Hilbert space for the $k$th pair, $k\in I_m=\{1,2,\ldots,m\}$. We assume that the Hilbert spaces $\pH_k^A$ and $\pH_k^B$ have the same finite dimension, which we denote by $d_k$. Although the case $d_k=1$ is not of interest for the distillability problem, we shall not exclude it for the sake of completness. The Hilbert space for the whole system of $m$ pairs of particles is $\pH=\pH^A\otimes\pH^B$, where
$$ \pH^A=\otimes_{k=1}^m \pH_k^A, \quad
\pH^B=\otimes_{k=1}^m \pH_k^B. $$
We refer to ${\bf d}=(d_1,\ldots,d_m)$ as the {\em dimension vector}
of this composite quantum system.
 
We fix an o.n. basis $\vek{i_k}$, $i_k=1,\ldots,d_k$ of $\pH_k^A$, and an o.n. basis of $\pH_k^B$ for which we use the same notation. Let $F_k:\pH_k\to\pH_k$ be the flip operator and 
$\rho_k^W(t)=1-tF_k$ the (non-normalized) Werner state. Let 
$\vek{\vf_k}\in\pH_k$ be the maximally entangled (pure) state 
and $P_k=\vek{\vf_k}\kov{\vf_k}$ its density matrix.

We assume that, for each $k$, the $k$th pair of particles shared by 
Alice and Bob is in some Werner state $\rho^W_k(t_k)$ with 
$t_k\le1/2$. We set ${\bf t}=t_1,\ldots,t_m$ and
$$ \rho_{\bf d}^W({\bf t})=
\rho_1^W(t_1)\otimes\rho_2^W(t_2)\otimes\cdots\otimes\rho_m^W(t_m). $$

By using the same argument as in the proof of \cite[Lemma 4]{DiV}
one can easily prove the following lemma.

\begin{lemma} \label{Lema-DiV}
If $\rho_{\bf d}^W({\bf t})$ is 1-indistillable and 
${\bf t}'=t'_1,\ldots,t'_m$ is such that $t'_i\le t_i$ for each
$i$, then $\rho_{\bf d}^W({\bf t}')$ is also 1-indistillable.
\end{lemma}

In view of this lemma, we assume from now on that all pairs of 
particles shared by Alice and Bob are in the critical Werner states 
$\rho_k^W=\rho_k^W(1/2)$, $1\le k\le m$, and we set
$$ \rho_{\bf d}^W=
\rho_1^W\otimes\rho_2^W\otimes\cdots\otimes\rho_m^W. $$
We shall refer to $\rho_{\bf d}^W$ as a {\em generalized critical 
Werner state}.

We are now ready to state our {\em Generalized Distillability Conjecture} (GDC).

\begin{conjecture} \label{GDC}
All generalized critical Werner states $\rho_{(d_1,\ldots,d_m)}^W$, 
$m\ge1$, are $1$-indistillable. 
\end{conjecture}

Note that since $m$ as well as the $d_k$ are arbitrary, if we replace ``1-indistillable'' with ``indistillable'' in GDC, we obtain an equivalent conjecture. In particular, GDC implies DC.

We note that GDC is valid in the case $m=1$ because the critical 
Werner states are 1-indistillable (see Prposition \ref{Werner-prop}).
A direct proof of a more general result will be given in Section
\ref{Specijalan}.

The partial transpose of $\rho_{\bf d}^W$ is 
\begin{equation} \label{parc-transp}
\sig_{\bf d}^W=(1-d_1 P_1/2)\otimes(1-d_2 P_2/2)\otimes\cdots
\otimes(1-d_m P_m/2).
\end{equation}

Hence, GDC can be restated as saying that the inequality
\begin{equation} \label{Nej-1}
\kov{\psi} \sig_{\bf d}^W \vek{\psi} \ge 0
\end{equation}
is valid for all $\vek{\psi}\in\pH$ of Schmidt rank at most two.

There is a conjecture \cite{LC,DiV,WD,PS} which asserts that 
the critical Werner states $\rho^W(1/2)$ are indistillable. 
Equivalently, it asserts that the tensor power 
$\rho^W(1/2)^{\otimes m}$ is $1$-indistillable for all $m\ge1$. 
This conjecture is a special case of GDC, which can be seen by 
setting all $d_k$ equal to $d$.

\section{Generalization of GDC and the CBS inequality}
 \label{Tenzor}
 
Our objective in this section is to reformulate GDC as an inequality 
for complex hypermatrices, and to embed this particular inequality 
in an infinite sequence of conjectural hypermatrix inequalities.

Our first job is to find an explicit expression for the inequality 
(\ref{Nej-1}) in terms of the components of $\vek{\psi}$. We have 
$\vek{\psi}=\vek{\psi_1}+\vek{\psi_2}$ where $\vek{\psi_1}$ and 
$\vek{\psi_2}$ are product vectors. Thus
\begin{eqnarray*} 
\vek{\psi_1} &=& \sum_{{\bf i},{\bf j}}
x_{i_1,\ldots,i_m}u_{j_1,\ldots,j_m}\vek{i_1,j_1,\ldots,i_m,j_m}, \\
\vek{\psi_2} &=& \sum_{{\bf i},{\bf j}}
y_{i_1,\ldots,i_m}v_{j_1,\ldots,j_m}\vek{i_1,j_1,\ldots,i_m,j_m},
\end{eqnarray*} 
where ${\bf i}=i_1,\ldots,i_m$; ${\bf j}=j_1,\ldots,j_m$
and, for each $s$, $i_s$ and $j_s$ run through the set $I_{d_s}$.

We can also write $\vek{\psi_1}=\vek{x}\otimes\vek{u}$ and $\vek{\psi_2}=\vek{y}\otimes\vek{v}$, where e.g.
$$ x=\sum_{\bf i} x_{i_1,\ldots,i_m}\vek{i_1,\ldots,i_m}\in\pH^A. $$
Thus $x=(x_{i_1,\ldots,i_m})$ is an $m$-dimensional complex matrix 
(or tensor with $m$ indexes). For simplicity, we shall refer to such
objects as {\em hypermatrices}. Let $\pM_{\bf d}$ denote the space of complex hypermatrices of type $d_1\times\cdots\times d_m$.
Formally they can be identified with the space of complex valued maps on the Cartesian product $I_{\bf d}=I_{d_1}\times\cdots\times I_{d_m}$. 
Finally we introduce the function $\Phi_{\bf d}:\pM_{\bf d}^4\to\bR$ 
by setting
$$ \Phi_{\bf d}(x,y,u,v)=\kov{\psi} \sig_{\bf d}^W \vek{\psi}. $$

We have to find explicit expressions for each of the four terms 
$\kov{\psi_i} \sig_{\bf d}^W \vek{\psi_j}$, $i,j\in\{1,2\}$. We shall 
derive the one for $i=1$ and $j=2$. Let also ${\bf p}=p_1,\ldots,p_m$ 
and ${\bf q}=q_1,\ldots,q_m$ where $p_s$ and $q_s$ will run through 
$I_{d_s}$. We start with
\begin{eqnarray*} 
\kov{\psi_1} \sig_{\bf d}^W \vek{\psi_2} &=& 
\sum_{{\bf i},{\bf j},{\bf p},{\bf q}}
x^*_{i_1,\ldots,i_m}u^*_{j_1,\ldots,j_m}
y_{p_1,\ldots,p_m}v_{q_1,\ldots,q_m} \cdot \\
&& \qquad\qquad \kov{i_1,j_1,\ldots,i_m,j_m} 
\sig_{\bf d}^W \vek{p_1,q_1,\ldots,p_m,q_m}
\end{eqnarray*} 

By expanding (\ref{parc-transp}) we obtain the formula
$$ \sig_{\bf d}^W=\sum_{z\in\{0,1\}^m} 
\left(\frac{-1}{2}\right)^{z_1+\cdots+z_m}
\sig_{z_1,1}\otimes\sig_{z_2,2}\otimes\cdots\otimes\sig_{z_m,m}, $$
where $z=(z_1,\ldots,z_m)$ and $\sig_{0,k}=1$, $\sig_{1,k}=d_kP_k$. 
Using this we find that
\begin{eqnarray*} 
\kov{\psi_1} \sig_{\bf d}^W \vek{\psi_2} &=& 
\sum_{Q\subseteq I_m} \left( \frac{-1}{2} \right)^{|Q|}
{\sum_{{\bf i},{\bf j},{\bf p},{\bf q}}}'
x^*_{i_1,\ldots,i_m}u^*_{j_1,\ldots,j_m}
y_{p_1,\ldots,p_m}v_{q_1,\ldots,q_m},
\end{eqnarray*} 
where $|Q|$ denotes the cardinality of $Q$ and
the summation in $\Sigma'$ is subject to the constraints
$$ p_s=i_s,\, q_s=j_s \textrm{ for } s\in I_m\setminus Q;\quad
j_s=i_s,\, q_s=p_s \textrm{ for } s\in Q. $$

For instance, if $m=2$ and $Q=\{2\}$ the constraints give $p_1=i_1$, $q_1=j_1$, $j_2=i_2$, $q_2=p_2$ and the sum $\Sigma'$ can be written as
$$ \sum_{i_1,j_1} \left(\sum_{i_2} x_{i_1,i_2}u_{j_1,i_2}\right)^*
\left(\sum_{p_2} y_{i_1,p_2}v_{j_1,p_2}\right). $$
In general, $\Sigma'$ can be written as 
$$ \sum_{i_r,j_r;\, r\in I_m\setminus Q} 
\left(\sum_{i_s,j_s=i_s;\, s\in Q} 
x_{i_1,\ldots,i_m}u_{j_1,\ldots,j_m}\right)^*
\left(\sum_{i_s,j_s=i_s;\, s\in Q} 
y_{i_1,\ldots,i_m}v_{j_1,\ldots,j_m}\right). $$
While performing the last two summations one has to set first 
$j_s=i_s$ for each $s\in Q$ and then to sum over all $i_s$, $s\in Q$.
Explicitly, using the Kronecker deltas, we have
$$ \sum_{i_s,j_s=i_s;\, s\in Q} y_{i_1,\ldots,i_m}v_{j_1,\ldots,j_m}=
\sum_{i_s,j_s;\, s\in Q} \quad \left( \prod_{q\in Q} \delta_{i_q,j_q}
\right) y_{i_1,\ldots,i_m}v_{j_1,\ldots,j_m}. $$
Similar three formulae are valid for 
$\kov{\psi_i}\sig_{\bf d}^W \vek{\psi_j}$, $(i,j)\ne(1,2)$. 
By using all four of them, we obtain that
\begin{equation} \label{Fi}
\Phi_{\bf d}(x,y,u,v) = 
\sum_{Q\subseteq I_m} \left( \frac{-1}{2} \right)^{|Q|}
\sum_{i_p,j_p;\, p\in I_m\setminus Q} 
\quad \left| \sum_{i_q,j_q=i_q;\, q\in Q} 
(x_{\bf i}u_{\bf j}+y_{\bf i}v_{\bf j}) \right|^2. 
\end{equation} 

Let us give two concrete examples. If $m=1$ and $d=d_1$, then
$$ \Phi_{(d)}(x,y,u,v)=\sum_{i,j=1}^d |x_iu_j+y_iv_j|^2
-\frac{1}{2}\left|\sum_{i=1}^d(x_iu_i+y_iv_i)\right|^2. $$
Writing $x,y,u,v$ as row vectors and using the Frobenius matrix 
norm, this can be expressed as
$$ \Phi_{(d)}(x,y,u,v)=\|x^Tu+y^Tv\|^2
-\frac{1}{2}\left|xu^T+yv^T\right|^2. $$

If $m=2$ then $x,y,u,v$ are ordinary $d_1\times d_2$ complex matrices and
\begin{eqnarray*} 
\Phi_{(d_1,d_2)}(x,y,u,v) &=& 
\sum_{i,j,k,l} |x_{ij}u_{kl}+y_{ij}v_{kl}|^2 -\frac{1}{2}\sum_{j,l} 
\left| \sum_i(x_{ij}u_{il}+y_{ij}v_{il}) \right|^2 \\
&& -\frac{1}{2}\sum_{i,k} \left|
\sum_j(x_{ij}u_{kj}+y_{ij}v_{kj})\right|^2  +\frac{1}{4}
\left| \sum_{i,j} (x_{ij}u_{ij}+y_{ij}v_{ij}) \right|^2,
\end{eqnarray*} 
where the indexes $i,k$ run through $I_{d_1}$ and $j,l$ through 
$I_{d_2}$. Using the tensor product of matrices, this can be 
written in a more compact form
\begin{eqnarray*}
\Phi_{(d_1,d_2)}(x,y,u,v) &=& 
\|x\otimes u+y\otimes v\|^2-\frac{1}{2} \| x^T u+y^T v \|^2 \\
&& -\frac{1}{2}\|ux^T+vy^T\|^2+\frac{1}{4}|\tr(x^T u+y^T v)|^2.
\end{eqnarray*}
In the special case $d_1=d_2=d$ this formula was derived in \cite{DZ1}.

{}From Eq. (\ref{Nej-1}) we see that GDC can be restated as follows. \begin{conjecture} \label{GDC-nej}
The inequality $\Phi_{\bf d}(x,y,u,v)\ge0$ is valid for all dimension 
vectors ${\bf d}=(d_1,\ldots,d_m)$, $m\ge1$, and all complex
hypermatrices $x,y,u,v\in \pM_{\bf d}$.
\end{conjecture}

We shall now embed this inequality in an infinite sequence of
hypermatrix inequalities. The sum 
$x_{\bf i}u_{\bf j}+y_{\bf i}v_{\bf j}$ occupies a prominent part 
in the definition of the function $\Phi_{\bf d}:\pM_{\bf d}^4\to\bR$ 
given by Eq. (\ref{Fi}). Clearly, it is rather unnatural for
this sum to have only two terms. Therefore we introduce a more 
general function $\Phi_{\bf d}^{(n)}:\pM_{\bf d}^{2n}\to\bR$ 
such that $\Phi_{\bf d}=\Phi_{\bf d}^{(2)}$. It is defined by a 
similar formula where the sum 
$x_{\bf i}u_{\bf j}+y_{\bf i}v_{\bf j}$ is replaced by one which has 
$n$ summands, and at the same time the fraction $-1/2$ is replaced 
with $-1/n$. As we will see in Section \ref{Osobine}, this new 
function shares with $\Phi_{\bf d}$ several properties observed 
in \cite{DZ1} in a special case, including the conjectural property 
of nonnegativity. The new function is defined as follows:
\begin{eqnarray}  \label{Gen-Fi} 
&& \Phi_{\bf d}^{(n)}(x^{(1)},\ldots,x^{(n)},
u^{(1)},\ldots,u^{(n)}) \\ \notag && \qquad 
= \sum_{Q\subseteq I_m} \left( \frac{-1}{n} \right)^{|Q|}
\sum_{i_p,j_p;\, p\in I_m\setminus Q} 
\quad \left| \sum_{i_q,j_q=i_q;\, q\in Q} \quad
\sum_{k=1}^n x_{\bf i}^{(k)}u_{\bf j}^{(k)} \right|^2,
\end{eqnarray} 
where ${\bf d}=(d_1,\ldots,d_m)$ and $x^{(k)},u^{(k)}\in\pM_{\bf d}$ 
are arbitrary hypermatrices.

We now state the {\em Cauchy-Bunyakovsky-Schwarz (CBS) conjecture}.
\begin{conjecture} \label{CBS-hip}
The inequality 
$$ \Phi_{\bf d}^{(n)}(x^{(1)},\ldots,x^{(n)},u^{(1)},\ldots,u^{(n)})\ge0 $$ 
is valid for all dimension vectors ${\bf d}=(d_1,\ldots,d_m)$, 
all $m,n\ge1$, and all complex hypermatrices 
$x^{(k)},u^{(k)}\in\pM_{\bf d}$.
\end{conjecture}

We will see later that this conjecture is true (and nontrivial) 
in the two boundary cases $m=1$ and $n=1$, see Sections 
\ref{Specijalan} and \ref{Lagrange} respectively. In the case $m=n=1$
this is indeed the classical Cauchy-Bunyakovsky-Schwarz inequality,
which explains our name for this conjecture. We also point out that 
the inequality in Conjecture \ref{GDC-nej} is just the case 
$n=2$ of the CBS conjecture. Thus the CBS conjecture generalizes 
both the GDC and the CBS inequality.

\section{Some properties of the function $\Phi_{\bf d}^{(n)}$} 
\label{Osobine}

We generalize here the properties of $\Phi_{\bf d}$
noted in \cite{DZ1} in a special case.

\subsection{Generalization of matrix transposition}
The transposition operation on ordinary matrices admits a wide  generalization to hypermatrices. Let us fix a permutation $\pi$ of the set $I_m=\{1,2,\ldots,m\}$. For a given dimension vector ${\bf d}=(d_1,\ldots,d_m)$ we set ${\bf d}'=(d'_1,\ldots,d'_m)$, where $d'_k=d_{\pi^{-1}k}$, $k\in I_m$. For $x\in \pM_{\bf d}$ we define 
$x'\in \pM_{{\bf d}'}$ by
$$ x'_{i'_1,\ldots,i'_m}=x_{i'_{\pi 1},\ldots,i'_{\pi m}}. $$
This is justified since $i'_k\in I_{d'_k}=I_{d_{\pi^{-1}k}}$ implies that $i'_{\pi k}\in I_{{\bf d}_k}$. Using the abbreviation 
${x'}^{(k)}=\left(x^{(k)}\right)'$, we have
\begin{eqnarray}  \label{Osobina1}
&& \quad \Phi_{{\bf d}'}^{(n)}({x'}^{(1)},\ldots,{x'}^{(n)},
{u'}^{(1)},\ldots,{u'}^{(n)}) \\ \notag && \qquad =  
\Phi_{\bf d}^{(n)}(x^{(1)},\ldots,x^{(n)},u^{(1)},\ldots,u^{(n)}) 
\end{eqnarray}
for all $x^{(k)},u^{(k)} \in \pM_{\bf d}$.

\subsection{Elimination of $d_s=1$}
We show that the $d_s$ which are equal to 1 can be eliminated. Assume that $d_s=1$ for some $s\in I_m$. Set ${\bf d}'=(d_1,\ldots,d_{s-1},d_{s+1},\ldots,d_m)$. For each $x\in\pM_{\bf d}$ we define 
$x'\in\pM_{{\bf d}'}$ by setting
$$ x'_{i_1,\ldots,i_{s-1},i_{s+1},\ldots,i_m}=
x_{i_1,\ldots,i_{s-1},1,i_{s+1},\ldots,i_m}. $$
Note that this $x'$ is different from the one used in the previous 
subsection.
\begin{proposition} \label{Stav-2}
Using the above assumption and the abbreviation 
${x'}^{(k)}=\left(x^{(k)}\right)'$, the equality
$$ \Phi_{\bf d}^{(n)}(x^{(1)},\ldots,x^{(n)},u^{(1)},\ldots,u^{(n)})
=\frac{n-1}{n} \Phi_{{\bf d}'}^{(n)}({x'}^{(1)},\ldots,{x'}^{(n)},
{u'}^{(1)},\ldots,{u'}^{(n)}) $$
is valid for any $x^{(k)},u^{(k)}\in\pM_{\bf d}$.
\end{proposition}

\noindent {\bf Proof.} 
Note that
\begin{equation} \label{Uproscena}
\Phi_{\bf d}^{(n)}(x^{(1)},\ldots,x^{(n)},u^{(1)},\ldots,u^{(n)})
=\sum_{Q\subseteq I_m}\left(\frac{-1}{n}\right)^{|Q|}\Phi_Q^{(n)},
\end{equation}
where 
\begin{eqnarray} \label{udeo}
&& \Phi_Q^{(n)}(x^{(1)},\ldots,x^{(n)},u^{(1)},\ldots,u^{(n)}) \\
\notag && \qquad\qquad = \sum_{i_p,j_p;\, p\in I_m\setminus Q}
\left| \sum_{i_q,j_q=i_q;\, q\in Q} 
\sum_{k=1}^n x_{\bf i}^{(k)}u_{\bf j}^{(k)} \right|^2.
\end{eqnarray}
We set $I'_m=I_m\setminus\{s\}$. Let $Q\subseteq I_m$ be a subset 
of cardinality $|Q|=k+1$ such that $s\in Q$, and set 
$Q'=Q\setminus\{s\}$. Because $d_s=1$ we have 
$\Phi_Q^{(n)}=\Phi_{Q'}^{(n)}$. Hence
$$ \left(\frac{-1}{n}\right)^{k+1} \Phi_Q^{(n)} 
+ \left(\frac{-1}{n}\right)^k \Phi_{Q'}^{(n)} = 
\frac{n-1}{n}\cdot \left(\frac{-1}{n}\right)^k \Phi_{Q'}^{(n)}. $$
It remains to observe that
\begin{eqnarray*}
&& \Phi_{{\bf d}'}^{(n)}
({x'}^{(1)},\ldots,{x'}^{(n)},{u'}^{(1)},\ldots,{u'}^{(n)}) 
\\ \notag && \qquad\qquad =
\sum_{Q'\subseteq I'_m} \left( \frac{-1}{n} \right)^{|Q'|}  
\Phi_{Q'}^{(n)}(x^{(1)},\ldots,x^{(n)},u^{(1)},\ldots,u^{(n)}).
\end{eqnarray*}
$\blacksquare$

We single out the following special case.

\begin{corollary} \label{Fi:n=1}
If ${\bf d}=(d_1,\ldots,d_m)$, $m\ge1$, and some $d_s=1$, then for
any $x,u\in\pM_{\bf d}$ we have $\Phi_{\bf d}^{(1)}(x,u)=0$.
\end{corollary}

\subsection{$\Un(d_s)$-invariance of $\Phi_{\bf d}^{(n)}$}
We introduce an action of the unitary group $\Un(d_s)$, 
$1\le s\le m$,
on the space $\pM_{\bf d}$. For $U\in\Un(d_s)$ and $x\in \pM_{\bf d}$
we define $U^{(s)}x\in \pM_{\bf d}$ by the formula
$$ (U^{(s)}x)_{i_1,\ldots,i_s,\ldots,i_m}=
\sum_{i'_s} U_{i_s,i'_s}x_{i_1,\ldots,i'_s,\ldots,i_m}. $$
To simplify the notation, we set $U^{*(s)}x={(U^*)}^{(s)}x$.

\begin{proposition} \label{Un-akcija}
$\Phi_{\bf d}^{(n)}$ is $\Un(d_s)$-invariant, i.e.,
\begin{eqnarray*} 
&& \Phi_{\bf d}^{(n)}(U^{(s)}x^{(1)},\ldots,U^{(s)}x^{(n)},
U^{*(s)}u^{(1)},\ldots,U^{*(s)}u^{(n)}) \\ && \qquad\qquad
=\Phi_{\bf d}^{(n)}(x^{(1)},\ldots,x^{(n)},u^{(1)},\ldots,u^{(n)})
\end{eqnarray*}
for all $x^{(k)},u^{(k)}\in \pM_{\bf d}$ and all $U\in\Un(d_s)$.
\end{proposition}

\noindent {\bf Proof.} 
In view of Eq. (\ref{Uproscena}), it suffices to show that for each 
$Q\subseteq I_m$ we have
\begin{eqnarray*}
&& \Phi_Q^{(n)}(U^{(s)}x^{(1)},\ldots,U^{(s)}x^{(n)},
U^{*(s)}u^{(1)},\ldots,U^{*(s)}u^{(n)}) \\ && \qquad\qquad
= \Phi_Q^{(n)}(x^{(1)},\ldots,x^{(n)},u^{(1)},\ldots,u^{(n)}),
\end{eqnarray*}
where $\Phi_Q^{(n)}$ is defined as in Eq. (\ref{udeo}).

If $s\in Q$ this is easy to verify. Indeed we have
\begin{eqnarray*}
\sum_{i_s,j_s=i_s}(U^{(s)}x)_{\bf i}(U^{*(s)}u)_{\bf j} &=& 
\sum_{i'_s,j'_s}\left(\sum_{i_s}U_{i_s,i'_s}U^*_{i_s,j'_s}\right)
x_{i_1,\ldots,i'_s,\ldots,i_m}u_{j_1,\ldots,j'_s,\ldots,j_m} \\
&=& \sum_{i'_s,j'_s}\delta_{i'_s,j'_s}
x_{i_1,\ldots,i'_s,\ldots,i_m}u_{j_1,\ldots,j'_s,\ldots,j_m} \\
&=& \sum_{i_s,j_s=i_s} x_{\bf i}u_{\bf j}.
\end{eqnarray*}

If $s\notin Q$ then we consider the sum
\begin{eqnarray*}
\sum_{i_s,j_s}\left|\sum_{i_q,j_q=i_q;q\in Q} \quad
\sum_{k=1}^n (U^{(s)}x^{(k)})_{\bf i} (U^{*(s)}u^{(k)})_{\bf j}
\right|^2 &=& \sum_{i_s,j_s}\left|
\sum_{i'_s,j'_s}U_{i_s,i'_s}z_{i'_s,j'_s}U^*_{j_s,j'_s} \right|^2 \\
&=& \|UZU^\dag\|^2,
\end{eqnarray*}
where $Z$ is the matrix with entries 
$$ z_{i'_s,j'_s}=\sum_{i_q,j_q=i_q;q\in Q} \quad
\sum_{k=1}^n x_{i_1,\ldots,i'_s,\ldots,i_m}^{(k)}
u_{j_1,\ldots,j'_s,\ldots,j_m}^{(k)};\quad i'_s,j'_s\in I_{d_s}. $$
Since $U$ is unitary we have 
$$ \|UZU^\dag\|^2 = \|Z\|^2 =\sum_{i_s,j_s}
\, \left| \sum_{i_q,j_q=i_q;\, q\in Q} \quad 
\sum_{k=1}^n x_{\bf i}^{(k)}u_{\bf j}^{(k)} \right|^2, $$
which completes the proof.
$\blacksquare$

The actions of the unitary groups $\Un(d_s)$ for $s\in I_m$ 
on $\pM_{\bf d}$ pairwise commute, and so we obtain an action of 
the direct product $\Un({\bf d})=\Un(d_1)\times\cdots\times\Un(d_m)$
on $\pM_{\bf d}$. Clearly, $\Phi_{\bf d}^{(n)}$ is also 
$\Un({\bf d})$-invariant.

\subsection{$\GL_n$-invariance of $\Phi_{\bf d}^{(n)}$}
Let $\La=[\al_{p,q}]\in\GL_n(\bC)$ and $(\La^T)^{-1}=[\be_{p,q}]$.

\begin{proposition} \label{GL2-akcija}
Using this notation, we have
\begin{eqnarray*}
&& \Phi_{\bf d}^{(n)} \left( \sum_{k=1}^n \al_{1,k}x^{(k)},\ldots,
\sum_{k=1}^n \al_{n,k}x^{(k)},\sum_{k=1}^n \be_{1,k}u^{(k)},\ldots,
\sum_{k=1}^n \be_{n,k}u^{(k)} \right) \\ && \qquad\qquad\qquad
= \Phi_{\bf d}^{(n)}(x^{(1)},\ldots,x^{(n)},u^{(1)},\ldots,u^{(n)})
\end{eqnarray*}
for all $x^{(k)},u^{(k)}\in\pM_{\bf d}$.
\end{proposition}

The proof given in \cite[Section 4]{DZ1} for a special case 
extends easily to the general case.

\section{Some special cases of the CBS conjecture} \label{Specijalan}

We start with a simple direct proof of the CBS conjecture in the case 
$m=1$. In particular, this gives an independent proof of GDC when 
$m=1$.

\begin{proposition} \label{CBS:m=1}
The CBS conjecture is true when $m=1$.
\end{proposition}
\noindent {\bf Proof.} 
We set $d_1=d$ and represent the hypermatrices 
$x^{(k)},u^{(k)}\in\pM_{(d)}$ as row vectors. The CBS conjecture 
takes the simple form:
$\|X\|^2\ge(1/n)|\tr X|^2$, where
$$ X=\left(x^{(1)}\right)^T u^{(1)}+\cdots+
\left(x^{(n)}\right)^T u^{(n)} $$
is a $d\times d$ matrix of rank $r\le n$. 
By Schur's triangularization theorem, we can choose $B\in\Un(d)$ 
such that $Y=BXB^\dag$ is upper triangular. Then the 
diagonal entries $\la_1,\ldots,\la_d$ of $Y$ are its eigenvalues 
and, by the same theorem, we may assume that $\la_k=0$ for 
$r<k\le d$. Since $\|X\|=\|Y\|$ and $\tr X=\tr Y$, our inequality
becomes $\|Y\|^2\ge(1/n)|\la_1+\cdots+\la_r|^2$. Since
$\|Y\|^2\ge|\la_1|^2+\cdots+|\la_r|^2$ and $r\le n$, the last 
inequality is a consequence of the well known inequality
$$ \frac{\mu_1^2+\cdots+\mu_r^2}{r}\ge
\left(\frac{\mu_1+\cdots+\mu_r}{r}\right)^2, $$
where $\mu_1,\ldots,\mu_r$ are nonnegative real numbers.
$\blacksquare$

In the case $n=2$ we can extend Proposition \ref{CBS:m=1} by making 
$m$  arbitrary provided that $d_i>2$ for at most one index $i$. 
However in this case we rely on the equivalence of GDC and 
the case $n=2$ of the CBS conjecture.

\begin{proposition} \label{Slucaj-2}
Let $\rho_i$, $i=1,2$, be a bipartite state acting on $\pH_i=\pH_i^A \otimes \pH_i^B$. If $\rho_1$ is separable and $\rho_2$ 
1-indistillable, then $\rho_1\otimes\rho_2$ is 1-indistillable.
Consequently, GDC is valid for arbitrary $m$ provided that 
$d_i>2$ for at most one index $i$. 
\end{proposition}
\noindent {\bf Proof.} 
Let $\sig_i=(T\otimes1)\rho_i$, $i=1,2$, be the partial transpose of $\rho_i$. Since $\rho_1$ is separable, $\sig_1$ is positive semidefinite and separable. Let $\vek{\psi}\in\pH_1\otimes\pH_2$ 
be any vector with Schmidt rank $\le2$ (with respect to the 
partition $A|B$). We have to show that 
$\kov{\psi} \sig_1\otimes\sig_2 \vek{\psi} \ge 0$. 
If $\tr_1$ is the first partial trace function (i.e., corresponding 
to $\pH_1$), then 
$$ \kov{\psi} \sig_1\otimes\sig_2 \vek{\psi}=
\tr\left(\vek{\psi}\kov{\psi} \sig_1\otimes\sig_2 \right)=\tr\left[
\tr_1\left(\vek{\psi}\kov{\psi}\sig_1\right)\sig_2)\right]. $$
We now use an argument from \cite[Section III B]{WD}. Since $\sig_1$ is separable, we can write it as 
$\sig_1=\sum_i c_i\vek{a_i,b_i}\kov{a_i,b_i}$, $c_i>0$. Then
$$ \tr_1\left(\vek{\psi}\kov{\psi}\sig_1\right)=
\sum_i c_i \vek{\psi_i}\kov{\psi_i}, $$
where $\vek{\psi_i}=\langle a_i,b_i|\psi\rangle$ is a state acting on $\pH_2$. Next we have
$$ \kov{\psi} \sig_1\otimes\sig_2 \vek{\psi}=
\sum_i c_i\kov{\psi_i} \sig_2 \vek{\psi_i}. $$
As the local projections cannot increase the Schmidt rank, each $\vek{\psi_i}$ has Schmidt rank $\le2$. Since $\rho_2$ is 1-indistillable, each summand in the above sum is nonnegative. This concludes the 
proof of the first assertion. 

The second follows from the first and the Eq. (\ref{Osobina1}).
$\blacksquare$

The proof of GDC can be reduced to the case where 
$3\le d_1\le d_2\le\cdots\le d_m$. This is the objective of the 
next proposition.

\begin{proposition}
If Conjecture \ref{GDC-nej} is true for all 
${\bf d}=(d_1,\ldots,d_m)$, $m\ge2$, satisfying 
$3\le d_1\le d_2\le\cdots\le d_m$, then it is true in general 
(i.e., without any restrictions on the dimension vector ${\bf d}$).
\end{proposition}

\noindent {\bf Proof.} 
Let ${\bf d}=(d_1,\ldots,d_m)$ be arbitrary and let
$\rho^W_{\bf d}$ be the corresponding generalized critical Werner 
state. We have to prove that $\Phi_{\bf d}(x,y,u,v)\ge0$ for all 
$x,y,u,v\in\pM_{\bf d}$. Because of Eq. (\ref{Osobina1}) we may 
assume that $d_1\le d_2\le\cdots\le d_m$. 
By Proposition \ref{Stav-2}, we can further assume that $2\le d_1$. 
If $3\le d_1$ then the assertion follows from the hypothesis. 

Otherwise we have $2=d_1=\cdots=d_k$ for some $k\le m$.
The bipartite state $\rho'=\rho^W_1\otimes\cdots\otimes\rho^W_k$ is
separable. If $k=m$ then $\rho^W_{\bf d}=\rho'$, and so it is 
separable and hence 1-indistillable. Finally, let $k<m$. Then the
bipartite state $\rho''=\rho^W_{k+1}\otimes\cdots\otimes\rho^W_m$ 
is 1-indistillable by hypothesis or (if $m=k+1$) by Proposition 
\ref{Werner-st}, part (c). Hence, $\rho^W_{\bf d}=\rho'\otimes\rho''$ 
is 1-indistillable by Proposition \ref{Slucaj-2}.
$\blacksquare$

\section{Hypermatrix Lagrange identities} \label{Lagrange}

It will be shown in this section that the CBS conjecture is true 
in the case $n=1$. Indeed, this is an immediate consequence of 
Theorem \ref{gen-Lagr-iden} below.

From Eq. (\ref{Gen-Fi}) we see that
$$ \Phi_{\bf d}^{(1)}(x,u)=\sum_{Q\subseteq I_m} (-1)^{|Q|}
\,\sum_{i_p,j_p;\,p\in I_m\setminus Q} \, \left| 
\sum_{i_q,j_q=i_q;\, q\in Q} x_{\bf i}u_{\bf j} \right|^2, $$
where ${\bf d}=(d_1,\ldots,d_m)$ and $x,u\in\pM_{\bf d}$. We shall 
prove that the inequalities $\Phi_{\bf d}^{(1)}(x,u)\ge0$ indeed 
hold.

We write in detail the first two cases. For $m=1$ we set $d=d_1$ 
and obtain
$$ \Phi_{(d)}^{(1)}(x,u)=\sum_{i=1}^d |x_i|^2 \cdot 
\sum_{j=1}^d |u_j|^2-\left| \sum_{k=1}^d x_k u_k \right|^2. $$
The hypermatrices $x$ and $u$ are just vectors in $\bC^d$.

For $m=2$ we have
\begin{eqnarray*}
\Phi_{(d_1,d_2)}^{(1)}(x,u) &=& \sum_{i_1,i_2} |x_{i_1,i_2}|^2 \cdot 
\sum_{j_1,j_2} |u_{j_1,j_2}|^2 -\sum_{i_2,j_2} \left| 
\sum_{i_1} x_{i_1,i_2} u_{i_1,j_2} \right|^2 \\ &&
-\sum_{i_1,j_1} \left| \sum_{i_2} x_{i_1,i_2} u_{j_1,i_2} \right|^2 
+\left| \sum_{i_1,i_2}  x_{i_1,i_2} u_{i_1,i_2} \right|^2, 
\end{eqnarray*}
where $i_1$ and $j_1$ run through $I_{d_1}$ and  $i_2$ and $j_2$ 
through $I_{d_2}$. In this case the hypermatrices $x$ and $u$ are 
ordinary $d_1\times d_2$ complex matrices.

The inequality $\Phi_{(d)}^{(1)}(x,u)\ge0$ is the well known
Cauchy-Bunyakovsky-Schwarz inequality, which follows trivially from 
the classical Lagrange identity
\begin{equation*}
\sum_{i=1}^d |x_i|^2 \cdot \sum_{j=1}^d |u_j|^2
-\left| \sum_{k=1}^d x_k u_k \right|^2 = 
\sum_{i,j;\, i<j} |x_i u_j^*-x_j u_i^* |^2.
\end{equation*} 

The inequality $\Phi_{(d_1,d_2)}^{(1)}(x,u)\ge0$ follows trivially 
from the following matrix analog of Lagrange identity:
\begin{eqnarray*} 
&& \Phi_{(d_1,d_2)}^{(1)}(x,u) = \\
&& \qquad \sum_{{\bf i},{\bf j};\,{\bf i}<{\bf j}} 
|x_{i_1,i_2} u_{j_1,j_2}^*-x_{i_1,j_2} u_{j_1,i_2}^*
-x_{j_1,i_2} u_{i_1,j_2}^*+x_{j_1,j_2} u_{i_1,i_2}^*|^2,
\end{eqnarray*} 
where we recall that ${\bf i}=i_1,i_2$ and ${\bf j}=j_1,j_2$,
and the inequality ${\bf i}<{\bf j}$ means that 
$i_1<j_1$ and $i_2<j_2$.

There is also an analog of Lagrange identity for any $m$. To state 
and prove these identities in general we need some additional notation. 
Let ${\bf i}=i_1,\ldots,i_m$ and ${\bf j}=j_1,\ldots,j_m$
be two sequences of indexes with $i_s,j_s\in I_{d_s}$
for each $s$. For arbitrary $Q\subseteq I_m$ we
define two modified sequences of indexes
${\bf i}^{Q,{\bf j}}$ and ${\bf j}^{Q,{\bf i}}$ of length $m$
as follows:
\begin{equation} \label{indeksi}
{\bf i}^{Q,{\bf j}}_p=\left\{
\begin{array}{ll} j_p & \mbox{if $p\in Q$;}\\ i_p & \mbox{otherwise;}
\end{array} \right. \quad
{\bf j}^{Q,{\bf i}}_p=\left\{
\begin{array}{ll} i_p & \mbox{if $p\in Q$;}\\ j_p & \mbox{otherwise.}
\end{array} \right.
\end{equation}
We also define
\begin{equation} \label{sigma-funk}
\sig_{{\bf i};{\bf j}}(x,u)=\sum_{Q\subseteq I_m}
(-1)^{|Q|} x_{{\bf i}^{Q,{\bf j}}} u_{{\bf j}^{Q,{\bf i}}}.
\end{equation}

For $m=1,2$ we have:
\begin{eqnarray*} 
\sig_{i;j}(x,u^*) &=& x_i u_j^*-x_j u_i^*, \\
\sig_{i_1,i_2;j_1,j_2}(x,u^*) &=& 
x_{i_1,i_2} u_{j_1,j_2}^*-x_{i_1,j_2} u_{j_1,i_2}^*
-x_{j_1,i_2} u_{i_1,j_2}^*+x_{j_1,j_2} u_{i_1,i_2}^*.
\end{eqnarray*} 

\begin{lemma} \label{znaci} 
Let ${\bf i}=i_1,\ldots,i_m$ and ${\bf j}=j_1,\ldots,j_m$ be two 
sequences of indexes with $i_s,j_s\in I_{d_s}$ for each $s$. 
For any $Q\subseteq I_m$ and any $x,u\in\pM_{\bf d}$ we have
$$ \sig_{{\bf i}';{\bf j}'}(x,u)=(-1)^{|Q|}
\sig_{{\bf i};{\bf j}}(x,u), $$
where ${\bf i}'={\bf i}^{Q,{\bf j}}$ and 
${\bf j}'={\bf j}^{Q,{\bf i}}$.
\end{lemma}
\noindent {\bf Proof.} 
We first observe that if $Q,R\subseteq I_m$ and 
\begin{eqnarray*}
&& {\bf i}'={\bf i}^{Q,{\bf j}},\quad {\bf j}'={\bf j}^{Q,{\bf i}};\\ 
&& {\bf i}''={{\bf i}'}^{R,{\bf j}'},\quad 
{\bf j}''={{\bf j}'}^{R,{\bf i}'};
\end{eqnarray*}
then
$$ {\bf i}''={\bf i}^{Q\triangle R,{\bf j}},\quad 
{\bf j}''={\bf j}^{Q\triangle R,{\bf i}}, $$
where $Q \triangle R$ is the symmetric difference of $Q$ and $R$.
Hence
\begin{eqnarray*}
\sig_{{\bf i}';{\bf j}'}(x,u) &=& \sum_{R\subseteq I_m}
(-1)^{|R|} x_{{{\bf i}'}^{R,{\bf j}'}}u_{{{\bf j}'}^{R,{\bf i}'}} \\
&=& \sum_{R\subseteq I_m}
(-1)^{|R|} x_{{\bf i}^{Q\triangle R,{\bf j}}} 
u_{{\bf j}^{Q\triangle R,{\bf i}}} \\
&=& (-1)^{|Q|} \sum_{R\subseteq I_m}
(-1)^{|Q\triangle R|} x_{{\bf i}^{Q\triangle R,{\bf j}}} 
u_{{\bf j}^{Q\triangle R,{\bf i}}} \\
&=& (-1)^{|Q|} \sig_{{\bf i};{\bf j}}(x,u),
\end{eqnarray*}
where we used the congruence 
$|Q\triangle R| \equiv |Q|+|R| \pmod{2}$ and the fact that 
when $R$ runs through all subsets of $I_m$ so does
$Q\triangle R$.
$\blacksquare$

In particular, note that if $i_s=j_s$ for some $s$ then 
$\sig_{{\bf i};{\bf j}}(x,u)=0$.

We can now state our hypermatrix Lagrange identity in full 
generality.
\begin{theorem} \label{gen-Lagr-iden}
For any dimension vector ${\bf d}=(d_1,\ldots,d_m)$, $m\ge1$, and any hypermatrices $x,u\in\pM_{\bf d}$ we have
\begin{equation}  \label{Lagr-m-jed}
\Phi_{\bf d}^{(1)}(x,u) =
\frac{1}{2^m} \sum_{{\bf i},{\bf j}} 
\left| \sig_{{\bf i};{\bf j}}(x,u^*) \right|^2 
= \sum_{{\bf i},{\bf j};\,{\bf i}<{\bf j}} \quad 
\left| \sig_{{\bf i};{\bf j}}(x,u^*) \right|^2,
\end{equation} 
where ${\bf i}<{\bf j}$ means that $i_s<j_s$ for all $s$.
\end{theorem}
\noindent {\bf Proof.} 
To prove the second equality, let $\Omega$ be the set of all 
pairs $({\bf i};{\bf j})$ such that $i_s\ne j_s$ for all $s$, and 
let $\Omega^\#$ be its subset consisting of all
pairs $({\bf i};{\bf j})$ such that $i_s<j_s$ for all $s$.
Since $\sig_{{\bf i};{\bf j}}(x,u^*)=0$ if $i_s=j_s$ for some $s$
and, by  Lemma \ref{znaci},
$$ \left| \sig_{{{\bf i}^{Q,{\bf j}}};{\bf j}^{Q,{\bf i}}}(x,u^*)
\right|^2 = \left| \sig_{{\bf i};{\bf j}}(x,u^*) \right|^2 $$
for all $Q\subseteq I_m$, we have
\begin{eqnarray*}
\sum_{{\bf i},{\bf j}}\left| \sig_{{\bf i};{\bf j}}(x,u^*) \right|^2 
&=&  \sum_{({\bf i};{\bf j}) \in \Omega} 
\left| \sig_{{\bf i};{\bf j}}(x,u^*) \right|^2 \\
&=&  \sum_{({\bf i};{\bf j}) \in \Omega^\#} \sum_{Q\subseteq I_m} 
\left| \sig_{{{\bf i}^{Q,{\bf j}}};{\bf j}^{Q,{\bf i}}}(x,u^*) 
\right|^2 \\
&=&  2^m \sum_{({\bf i};{\bf j}) \in \Omega^\#}
\left| \sig_{{\bf i};{\bf j}}(x,u^*) \right|^2, \\
&=& 2^m \sum_{{\bf i},{\bf j};\,{\bf i}<{\bf j}} \quad 
\left| \sig_{{\bf i};{\bf j}}(x,u^*) \right|^2.
\end{eqnarray*}

The proof of the first equality is more involved. From Eq. 
(\ref{sigma-funk}) we obtain that
$$ \left| \sig_{{\bf i};{\bf j}}(x,u^*) \right|^2=
\sum_{Q,R\subseteq I_m} (-1)^{|Q \triangle R|}
x_{{\bf i}^{Q,{\bf j}}} u_{{\bf j}^{R,{\bf i}}}
x^*_{{\bf i}^{R,{\bf j}}} u^*_{{\bf j}^{Q,{\bf i}}}. $$
Hence
\begin{equation} \label{zbir-1}
\sum_{{\bf i},{\bf j}}\left|\sig_{{\bf i};{\bf j}}(x,u^*)\right|^2=
\sum_{Q,R\subseteq I_m} (-1)^{|Q \triangle R|}
\sum_{{\bf i},{\bf j}} 
x_{{\bf i}^{Q,{\bf j}}} u_{{\bf j}^{R,{\bf i}}}
x^*_{{\bf i}^{R,{\bf j}}} u^*_{{\bf j}^{Q,{\bf i}}}.
\end{equation}

{}From Eq. (\ref{indeksi}) we see that, for a fixed $s\in I_m$, 
we have
\begin{equation*}
\left({\bf i}_s^{Q,{\bf j}},{\bf j}_s^{R,{\bf i}},
{\bf i}_s^{R,{\bf j}},{\bf j}_s^{Q,{\bf i}}\right)=\left\{
\begin{array}{ll}
(i_s,i_s,j_s,j_s)& \mbox{if $s\in R\setminus Q$;} \\
(j_s,j_s,i_s,i_s)& \mbox{if $s\in Q\setminus R$;} \\
(j_s,i_s,j_s,i_s)& \mbox{if $s\in Q\cap R$;} \\
(i_s,j_s,i_s,j_s)& {\rm otherwise.}
\end{array} \right.
\end{equation*}
By using this, it is not hard to deduce that
\begin{equation} \label{zbir-2}
\sum_{{\bf i},{\bf j}} 
x_{{\bf i}^{Q,{\bf j}}} u_{{\bf j}^{R,{\bf i}}}
x^*_{{\bf i}^{R,{\bf j}}} u^*_{{\bf j}^{Q,{\bf i}}}=
\sum_{i_s,j_s;\, s\notin Q \triangle R}
\left| \sum_{i_s,j_s=i_s;\, s\in Q \triangle R}
x_{{\bf i}^{Q,{\bf j}}} u_{{\bf j}^{R,{\bf i}}} \right|^2.
\end{equation} 

The correct interpretation of this equality requires that we
omit the summation sign on the right-hand side if the set of 
indexes over which we sum is empty. For instance let $m=2$ and 
$Q=R=\{1\}$. Then 
${\bf i}^{Q,{\bf j}}={\bf i}^{R,{\bf j}}=j_1,i_2$ and
${\bf j}^{Q,{\bf i}}={\bf j}^{R,{\bf i}}=i_1,j_2$, and we have
\begin{eqnarray*}
\sum_{{\bf i},{\bf j}} 
x_{{\bf i}^{Q,{\bf j}}} u_{{\bf j}^{R,{\bf i}}}
x^*_{{\bf i}^{R,{\bf j}}} u^*_{{\bf j}^{Q,{\bf i}}}
&& = \sum_{{\bf i},{\bf j}} x_{j_1,i_2}u_{i_1,j_2}
x^*_{j_1,i_2}u^*_{i_1,j_2} \\
&& = \sum_{{\bf i},{\bf j}} |x_{j_1,i_2}u_{i_1,j_2}|^2.
\end{eqnarray*}
This agrees with Eq. (\ref{zbir-2}) provided that
$$ \sum_{i_s,j_s=i_s;\, s\in Q \triangle R}
x_{{\bf i}^{Q,{\bf j}}} u_{{\bf j}^{R,{\bf i}}} =
x_{{\bf i}^{Q,{\bf j}}} u_{{\bf j}^{R,{\bf i}}} \left( =
x_{j_1,i_2}u_{i_1,j_2} \right). $$
According to our interpretation, this equality is indeed valid
since $Q \triangle R=\emptyset$.

Since for a fixed $S\subseteq I_m$ there are exactly $2^m$ choices 
for subsets $Q,R\subseteq I_m$ such that $Q \triangle R=S$, we
deduce from Eqs. (\ref{zbir-1}) and (\ref{zbir-2}) that
\begin{equation*}
\sum_{{\bf i},{\bf j}}\left|\sig_{{\bf i};{\bf j}}(x,u^*)\right|^2=
2^m \sum_{S\subseteq I_m} (-1)^{|S|} 
\sum_{i_s,j_s;\, s\in I_m\setminus S} \left|
\sum_{i_s,j_s=i_s;\, s\in S} x_{\bf i} u_{\bf j} \right|^2.
\end{equation*} 
This completes the proof of the theorem.
$\blacksquare$

Observe that if some $d_s=1$ then the inequality ${\bf i}<{\bf j}$
cannot be satisfied and so Eq. (\ref{Lagr-m-jed}) gives
$\Phi_{\bf d}^{(1)}(x,u)=0$. This agrees with Corollary \ref{Fi:n=1}.

The following purely algebraic version of the generalized Lagrange 
identity is valid in a more general context.

\begin{corollary} \label{Lagr-ident}
For any dimension vector ${\bf d}=(d_1,\ldots,d_m)$, $m\ge1$, and any hypermatrices $x,u$ of type $d_1\times\cdots\times d_m$ with values
in an arbitrary commutative ring, we have
\begin{equation} \label{Lagr-alg}
\sum_{Q\subseteq I_m} (-1)^{|Q|}
\,\sum_{i_p,j_p;\,p\in I_m\setminus Q} \, \left( \sum_{i_q;\, q\in Q} 
x_{{\bf i}^{Q,{\bf j}}}u_{{\bf j}^{Q,{\bf i}}} \right)^2
= \sum_{{\bf i},{\bf j};\,{\bf i}<{\bf j}} \,
\sig_{{\bf i};{\bf j}}(x,u)^2,
\end{equation} 
where we make use of the definitions in 
Eqs. (\ref{indeksi}) and (\ref{sigma-funk}).
\end{corollary}

\noindent {\bf Proof.} 
It is immediate from the theorem that the identity 
Eq. (\ref{Lagr-alg}) is valid when the hypermatrices $x,u$ are 
real-valued. Since this is a polynomial identity over the ring of 
integers, it is also valid when $x$ and $u$ are hypermatrices 
taking values in an arbitrary commutative ring.
$\blacksquare$

\section{Integral version of the CBS conjecture} \label{Integral}

Let $L^2(\bR^m)$ be the space of complex-valued square-integrable 
functions $\xi:\bR^m\to\bC$. We define the function
$\Phi_m^{(n)}:L^2(\bR^m)^{2n}\to\bR$ by the formula
\begin{eqnarray*}
&& \Phi_m^{(n)}(\xi^{(1)},\ldots,\xi^{(n)},
\eta^{(1)},\ldots,\eta^{(n)}) \\ 
&& = \sum_{Q\subseteq I_m} \left(\frac{-1}{n}\right)^{|Q|}
\int\cdots\int_{s_p,t_p;\, p\in I_m\setminus Q}
\left| \int\cdots\int_{s_q,t_q=s_q;\, q\in Q} \,\,
\sum_{k=1}^n \xi^{(k)}({\bf s})\eta^{(k)}({\bf t}) \right|^2,
\end{eqnarray*}
where $\xi^{(k)},\eta^{(k)}\in L^2(\bR^m)$ and
${\bf s}=s_1,\ldots,s_m$ and ${\bf t}=t_1,\ldots,t_m$. 
While computing the inner integrals (those within the
absolute value signs) one has first to set $t_q=s_q$ 
for each $q\in Q$ and then to integrate over each variable $s_q$, 
$q\in Q$. The outer integrations are performed over each of the 
variables $s_p$ and $t_p$ with $p\in I_m\setminus Q$.

In the case when the functions  $\xi^{(k)},\eta^{(k)}$ are 
continuous with compact support, one can approximate them
by hypermatrices $x^{(k)},u^{(k)}\in\pM_{\bf d}$ where
${\bf d}=(d_1,\ldots,d_m)$. After suitable scaling and by
taking the limit as all $d_s\to\infty$, we obtain a version of 
the CBS conjecture in which all the sums are replaced by integrals. 
Since the continuous functions with compact support are dense
in $L^2(\bR^m)$, the CBS conjecture implies that 
\begin{equation}  \label{CBS-int-1}
\Phi_m^{(n)}(\xi^{(1)},\ldots,\xi^{(n)},
\eta^{(1)},\ldots,\eta^{(n)}) \ge0,
\end{equation} 
for arbitrary $\xi^{(k)},\eta^{(k)}\in L^2(\bR^m)$.

For instance if $m=n=2$ then the inequality reads as follows:
\begin{eqnarray*}  
\int\int\int\int\left| 
\xi^{(1)}(s_1,s_2) \eta^{(1)}(t_1,t_2)+
\xi^{(2)}(s_1,s_2) \eta^{(2)}(t_1,t_2) 
\right|^2 ds_1\;ds_2\;dt_1\;dt_2 \\
-\frac{1}{2}\int\int\left|\int\left(
\xi^{(1)}(s_1,s_2) \eta^{(1)}(s_1,t_2)+
\xi^{(2)}(s_1,s_2) \eta^{(2)}(s_1,t_2) 
\right) ds_1 \right|^2 ds_2\;dt_2 \\
-\frac{1}{2}\int\int\left|\int\left(
\xi^{(1)}(s_1,s_2) \eta^{(1)}(t_1,s_2)+
\xi^{(2)}(s_1,s_2) \eta^{(2)}(t_1,s_2) 
\right) ds_2 \right|^2 ds_1\;dt_1 \\
+\frac{1}{4}\left|\int\int\left(
\xi^{(1)}(s_1,s_2) \eta^{(1)}(s_1,s_2)+
\xi^{(2)}(s_1,s_2) \eta^{(2)}(s_1,s_2)
\right)ds_1\;ds_2\right|^2 \ge0,
\end{eqnarray*}
where each integral is taken over the whole real line. The four
expressions on the left-hand side are contributions of the four
subsets $Q=\emptyset,\{1\},\{2\},\{1,2\}$ of $I_2$, respectively.
However, we want to point out that this particular inequality is
only conjectural.

Since we know that the CBS conjecture is true if $m=1$ or $n=1$,
the inequality (\ref{CBS-int-1}) is valid in these two cases.
Explicitly, these inequalities are:
\begin{eqnarray*}
&& \Phi_1^{(n)}(\xi^{(1)},\ldots,\xi^{(n)},
\eta^{(1)},\ldots,\eta^{(n)})= \\ && \quad 
\int\int\left|\sum_{k=1}^n \xi^{(k)}(s)\eta^{(k)}(t) 
\right|^2 ds\;dt-\frac{1}{n}\left| \int
\sum_{k=1}^n \xi^{(k)}(s)\eta^{(k)}(s) ds \right|^2 \ge0
\end{eqnarray*}
and
\begin{eqnarray*}
&& \Phi_m^{(1)}(\xi,\eta)= \\
&& \sum_{Q\subseteq I_m} (-1)^{|Q|}
\int\cdots\int_{s_p,t_p;\, p\in I_m\setminus Q}\left|
\int\cdots\int_{s_q,t_q=s_q;\,q\in Q} \xi({\bf s})\eta({\bf t})
\right|^2\ge0.
\end{eqnarray*}

The latter inequality is also a consequence of the integral 
version of Eq. (\ref{Lagr-m-jed}) which we are going to write 
down explicitly. Recall that ${\bf s}=s_1,\ldots,s_m$ and 
${\bf t}=t_1,\ldots,t_m$ are sequences of real variables. 
For $Q\subseteq I_m$ we define two modified sequences, 
${\bf s}^{Q,{\bf t}}$ and ${\bf t}^{Q,{\bf s}}$, 
of the same length $m$:
$$
{\bf s}^{Q,{\bf t}}_p=\left\{
\begin{array}{ll} t_p & \mbox{if $p\in Q$;}\\ s_p & \mbox{otherwise;}
\end{array} \right. \quad
{\bf t}^{Q,{\bf s}}_p=\left\{
\begin{array}{ll} s_p & \mbox{if $p\in Q$;}\\ t_p & \mbox{otherwise.}
\end{array} \right.
$$
We also define
$$ \sig_{{\bf s};{\bf t}}(\xi,\eta)=\sum_{Q\subseteq I_m}
(-1)^{|Q|} \xi({\bf s}^{Q,{\bf t}}) \eta({\bf t}^{Q,{\bf s}}). $$

Now the integral version of the hypermatrix Lagrange identity 
(\ref{Lagr-m-jed}) can be stated as follows:
\begin{eqnarray*}  
\Phi_m^{(1)}(\xi,\eta) &=& \frac{1}{2^m} 
\int\cdots\int_{s_p,t_p;\, p\in I_m} 
\left| \sig_{{\bf s};{\bf t}}(\xi,\eta^*) \right|^2 \\
&=& \int\cdots\int_{s_p<t_p;\, p\in I_m} 
\left| \sig_{{\bf s};{\bf t}}(\xi,\eta^*) \right|^2,
\end{eqnarray*}
where $\xi,\eta\in L^2(\bR^m)$ are arbitrary.

We can derive additional inequalities from the CBS conjecture by
letting only some of the dimensions $d_s\to\infty$ and keeping
the other fixed. As an example we consider the case where
$m=n=2$ and ${\bf d}=(d_1,d_2)$, where $d_1=2$ while we let 
$d_2\to\infty$. By Proposition \ref{Slucaj-2}, we have 
$\Phi_{\bf d}(x^{(1)},x^{(2)},u^{(1)},u^{(2)})\ge0$ for all
$x^{(1)},x^{(2)},u^{(1)},u^{(2)}\in\pM_{\bf d}$. In this case 
we obtain the inequality
\begin{eqnarray*}  
&& \sum_{i,j}\int\int\left|
\xi^{(1)}_i(s) \eta^{(1)}_j(t)+
\xi^{(2)}_i(s) \eta^{(2)}_j(t)
\right|^2 ds\;dt \\
&& \quad -\frac{1}{2}\int\int\left|\sum_i\left(
\xi^{(1)}_i(s) \eta^{(1)}_i(t)+
\xi^{(2)}_i(s) \eta^{(2)}_i(t)
\right) \right|^2 ds\;dt \\
&& \quad -\frac{1}{2}\sum_{i,j}\left|\int\left(
\xi^{(1)}_i(s) \eta^{(1)}_j(s)+
\xi^{(2)}_i(s) \eta^{(2)}_j(s)
\right) ds \right|^2 \\
&& \quad +\frac{1}{4}\left|\sum_i\int\left(
\xi^{(1)}_i(s) \eta^{(1)}_i(s)+
\xi^{(2)}_i(s) \eta^{(2)}_i(s)
\right) ds \right|^2 \ge0,
\end{eqnarray*}
valid for arbitrary functions 
$\xi^{(k)}_i,\eta^{(k)}_j\in L^2(\bR)$, $k=1,2$. 
(As $d_1=2$, the indexes $i$ and $j$ run through $I_2=\{1,2\}$.)

We give two concrete examples.

First, let $\xi_i^{(k)}(t)=t^{(a_{ik}-1)/2}$ and 
$\eta_i^{(k)}(t)=t^{(b_{ik}-1)/2}$ where $0<t<1$ and all 
$a_{ik}$ and $b_{ik}$ are positive.
We assume that all of these functions vanish for $t\le0$ and 
for $t\ge1$. Then the above inequality becomes
\begin{eqnarray*}  
&& 4\sum_{i,j,k,l} \frac{1}{(a_{ik}+a_{il})(b_{jk}+b_{jl})}
-2\sum_{i,j,k,l} \frac{1}{(a_{ik}+a_{jl})(b_{ik}+b_{jl})} \\
&& \qquad -2\sum_{i,j} \left(\sum_k\frac{1}{a_{ik}+b_{jk}}
\right)^2+\left( \sum_{i,k} \frac{1}{a_{ik}+b_{ik}} \right)^2 \ge 0,
\end{eqnarray*}
where $i,j,k,l$ run independently through $\{1,2\}$.

Second, let $\xi_i^{(k)}(t)=e^{-a_{ik}t^2}$ and 
$\eta_i^{(k)}(t)=e^{-b_{ik}t^2}$ where all 
$a_{ik}$ and $b_{ik}$ are positive. Then we obtain the inequality 
\begin{eqnarray*}  
&& 4\sum_{i,j,k,l} \frac{1}{\sqrt{(a_{ik}+a_{il})(b_{jk}+b_{jl})}}
-2\sum_{i,j,k,l} \frac{1}{\sqrt{(a_{ik}+a_{jl})(b_{ik}+b_{jl})}} \\
&& \qquad -2\sum_{i,j} \left(\sum_k\frac{1}{\sqrt{a_{ik}+b_{jk}}}
\right)^2+\left( \sum_{i,k} \frac{1}{\sqrt{a_{ik}+b_{ik}}} 
\right)^2 \ge 0.
\end{eqnarray*}

\section{Conclusion} \label{Zakljucak}

The Distillation Conjecture (DC) asserts that there exist 
indistillable NPT (necessarily entangled) states in a bipartite 
$d\times d$ quantum system with $d\ge3$. Although it is more than 
ten years old, it remains open for all $d\ge3$.
DC has been reduced to the case of non-normalized Werner states 
$\rho^W(t)=1-tF$, $-1\le t\le1$, where $F$ is the flip operator.
For the affirmative answer to DC, it suffices to show 
(and it has been conjectured) that the critical Werner states 
$\rho^W=\rho^W(1/2)$ are indistillable. These critical Werner 
states are known to be 1-indistillable and lie on the boundary 
between the 1-distillable and the 1-indistillable Werner states.

We propose much stronger Generalized Distillation Conjecture (GDC) 
which asserts that the tensor product of arbitrary critical
Werner states $\rho_k^W$ of type $d_k\times d_k$, $k=1,\ldots,m$, 
is 1-indistillable. It is easy to see that ``1-indistillable'' 
can be replaced here by ``indistillable''.
We recall from \cite{PS} that there is a distillable bipartite 
state $\rho_{\bf Pyr}\otimes\rho^W(1/2)$, where $\rho_{\bf Pyr}$ 
is a particular PPT entangled state of two qutrits and 
$\rho^W(1/2)$ the critical Werner state of two qutrits. 
In view of this example, GDC appears to be somewhat 
counter-intuitive as it asserts in particular that 
$\rho^W(1/2)\otimes\rho^W(1/2)$ is indistillable. 
However, there is a strong numerical evidence that the latter 
state is indistillable.

We reformulate the GDC as an inequality
$\Phi_{\bf d}(x,y,u,v)\ge0$, see Conjecture \ref{GDC-nej}, where 
${\bf d}=(d_1,\ldots,d_m)$ and $x,y,u,v$ are arbitrary complex 
hypermatrices of type $d_1\times\cdots\times d_m$. 
We are then led to propose a more general inequality 
$\Phi_{\bf d}^{(n)}(x^{(1)},\cdots,x^{(n)},
u^{(1)},\cdots,u^{(n)})\ge0$, see Conjecture \ref{CBS-hip},
which includes GDC as the special case $n=2$. It also includes 
the classical Cauchy-Bunyakovsky-Schwarz (CBS) inequality as
the case $m=n=1$. For that reason we refer to the latter 
conjecture as the CBS conjecture. 

We have shown that the CBS conjecture is true in the two boundary 
cases $m=1$ and $n=1$. The case $n=1$ is quite interesting since 
we were able to express the function $\Phi_{\bf d}^{(1)}(x,u)$ 
explicitly as the sum of squares of real-valued polynomials. 
These polynomial identities provide generalizations of the 
classical Lagrange identity, which is obtained for $m=1$.

As in the well-known case of the classical CBS inequality,
the discrete inequalities in the CBS conjecture have their 
continous counterparts in which the hypermatrices are replaced with
complex-valued square-integrable functions on $\bR^m$, and 
the summation operations are replaced with suitable intagrals. 
We formulate explicitly these integral inequalities and point out 
some special cases including two concrete examples. We also give 
the integral version of the generalized Lagrange identities.

We hope that our conjectures and results will stimulate further 
research on the original Distillation Conjecture.

\end{document}